\documentclass[usenatbib]{mn2e}
\usepackage{psfig}
\newcommand{\ltaraw}{$\; \buildrel < \over \sim \;$}
\newcommand{\lta}{\lower.5ex\hbox{\ltaraw}}
\newcommand{\gtaraw}{$\; \buildrel > \over \sim \;$}
\newcommand{\gta}{\lower.5ex\hbox{\gtaraw}}

\def\ltsima{$\; \buildrel < \over \sim \;$}
\def\simlt{\lower.5ex\hbox{\ltsima}}
\def\gtsima{$\; \buildrel > \over \sim \;$}
\def\simgt{\lower.5ex\hbox{\gtsima}}

\newcommand{\kms}{{\rm\,km\,s^{-1}}}
\newcommand{\kpc}{{\rm\,kpc}}

\loadboldmathitalic 
\title [LMC self-lensing]
{Kinematic outliers in the LMC: constraints on star-star microlensing}
\author[Zhao, Ibata, Lewis, Irwin]
{HongSheng Zhao$^1$, 
Rodrigo A. Ibata$^2$,
Geraint F. Lewis$^{3,4}$ \&
Michael J. Irwin$^1$ \\ 
$^1$ Institute of Astronomy, Madingley Road, Cambridge, CB3 0HA, U.K.:
Email \tt{zhao,mike@ast.cam.ac.uk} \\
$^2$
Observatoire de Strabourg, 11, rue de l'Universite, F-67000, Strasbourg, 
France:
Email \tt{ibata@astro.u-strasbg.fr} \\
$^3$Anglo-Australian Observatory, P.O. Box 296, Epping, NSW 1710, Australia:
Email \tt{gfl@aaoepp.aao.gov.au}\\
$^4$Present Address:
School of Physics, University of Sydney, NSW 2006, Australia:
Email {\tt gfl@physics.usyd,edu.au}
}
\date{MNRAS, in press (2003)}
\begin{document} 
\maketitle 
\begin{abstract}
Although  a  decade  of  microlensing  searches towards  
the Large Magellanic Cloud (LMC)  has
detected  $13-25$ possible  microlensing  events, the  nature and  the
location of the lenses, being either halo machos or LMC stars, remains
a subject of debate.  The star-star lensing models generically predict
the existence  of a small population  (more than $\sim  5$\%) of stars
with  a spatial and  kinematic distribution  different from  the thin,
young  disc of  the  LMC.  Here  we  present the  results  of a  large
spectroscopic survey of  the LMC, consisting of more  than 1300 radial
velocities measured accurately with the 2dF instrument.  In this large
sample, no  evidence is found  for any extraneous population  over the
expected LMC  and Galactic components.   Any additional, kinematically
distint, population  can only be present  at less than  the 1\% level.
We discuss the  significance of this finding for  the LMC self-lensing
models.
\end{abstract}

\section{Introduction}

About a decade of searches for  microlenses along the line of sight to
the  Magellanic Clouds (MCs) 
by  MACHO/OGLE/EROS and  a number  of follow-up  surveys have
found more than $13$ candidate microlensing events towards the LMC and
two  towards  the   SMC.   The  MACHO  collaboration  \citep{alcock00}
conclude that the lenses are mostly machos residing in the halo of the
Galaxy, perhaps in the form  of ancient white dwarfs (WDs).  These WDs
would make up  10-20\% of the halo dynamical  mass, with the remaining
80-90\% in some other component.   The detection of high proper motion
white  dwarfs  (\citealt{hambly97, hodgkin00, ibata00, oppenheimer01})
supports  this interpretation.

The assertion that WDs make up at least 3\% of the expected local dark
matter   density  \citet{oppenheimer01}   has  been   hotly  contested
\cite[see review by][]{richer01}.  Some argue very strongly that these
WDs  trace   the  thick  disk   kinematics  instead  (\citealt{reid01}).  
If the thick disk interpretation is true, then the halo
would have  very few WDs, but  the big question  remains: what objects
are responsible for microlensing the LMC stars?

The  answer  is perhaps  the  LMC  stars  themselves.  From  the  very
beginning there  have been  many variations of  star-star self-lensing
proposals  as first advocated  by \citet{sahu94}.   
Star-star  lensing models typically  invoke an 
unvirialised component of the LMC and the SMC 
(e.g., a puffed up disc or a  hot stellar halo with a surface brightness only
a few percent of  the LMC disc, a wrapped-around tidal ring, an offset bar 
etc, all due to tidal shocking among the LMC, SMC and the Galaxy) because
of the low self-lensing optical depth of a thin virialised LMC disk (Gould 1995).  
Any unvirialised component of the LMC can then lense with stars in the 
thin disc of the LMC.  Some recent theoretical models 
\citep[ e.g.,]{zhao01, jetzer02} argue that  there  could be enough  stellar
lenses in the LMC bar and disk to  account for from half to all of the observed
events.  These models generally predict that the  events should have
peculiarities  in  photometry,  kinematics and  spatial  distribution
because the lensing optical depth is higher in some regions than the others.
These are highlighted  in at least five of  the dozen observed events:
the LMC near-clump event MACHO-LMC-1 and LMC binary events MACHO-LMC-9 and
MACHO-LMC-14, the  SMC caustic event MACHO-98-SMC-1  and long duration
event MACHO-97-SMC-1.  
Although there are no strong direct evidences, it is reasonable 
to expect that all of the dozen observed lenses belong to the
same LMC or SMC population as for these exotic events
because ordinary single stars should be at least
as common as binary stars in a given LMC or SMC stellar population, 
and they all have similar lensing cross-sections and detection efficiencies.

\citet{zhao99a, zhao99b}  discussed a number of observational
tests  to differentiate the  two competing  classes of  lensing model.
In particular, stars  in any proposed  unvirialised component should
be identifiable  as kinematic outliers, meaning  that they
deviate from the rotation curve of  the LMC, as traced by the majority
of  LMC disc  stars  and by  HI  gas.  The  objective  of the  present
contribution is  to analyse  the kinematics of  large sample  of stars
towards the  LMC to investigate  whether outliers can be  singled out.
These kinematic outliers may reveal  the presence of any hot component
around the LMC, or any cold halo stream either in the Galactic halo or
the LMC  halo (Graff et al. 2000). We aim  to be able  to detect a  kinematically distinct
population (polar  ring or thickened disk)  in the LMC  present at the
$\sim 5$\% level.   If such a population is  identified their relation
to the microlensed sources can be studied.

We foresaw two possible outcomes of this experiment:
\begin{enumerate}
\item
The LMC microlensed sources and  the randomly selected LMC field stars
have  disk kinematics: This  would imply  that the  lenses are  in the
halo, supporting the MACHO interpretation of a dark baryonic component
in  the  halo in  the  form  of white  dwarfs  and  consistent 
with the high-velocity, blue,  old white dwarfs  
in the solar  neighbourhood detected by 
Ibata et al. (2000) and Oppenheimer et al. (2000).
\item
The sample  of microlenses or the  sample of field  stars contain many
outliers from the LMC rotation  curve, implying that the Galactic dark
halo is  almost entirely made  of a non-compact component.   The solar
neighbourhood white dwarfs are merely a local peculiarity.
\end{enumerate}

\begin{figure}
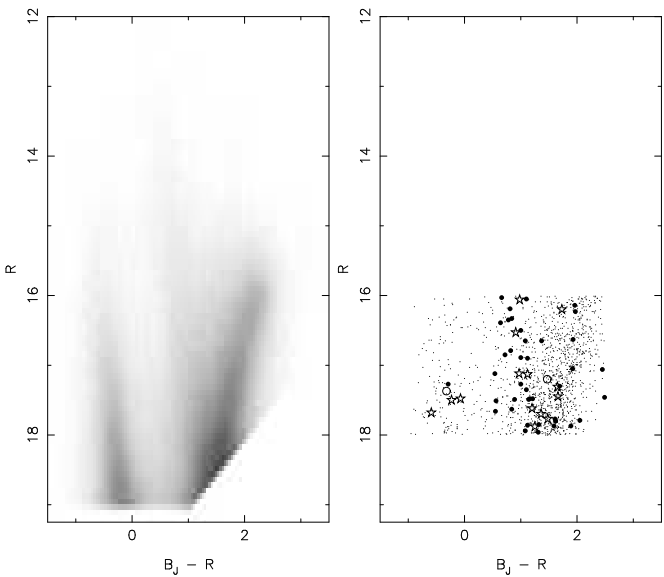

\hbox{
\psfig{figure=MC964fig1a.eps,angle=0,width=1.7in}
\psfig{figure=MC964fig1b.eps,angle=0,width=1.7in}}
\caption[]{The left-hand panel shows a  Hess diagram of the LMC field.
The right-hand panel  shows the portion of CMD  where our sample stars
are selected, from a box  of $16<R<18$ and $-1.0<B_{\rm J}-R<2.5$. The
targets are marked  according to the radial velocity  measured in this
survey: stars  with $-100\kms < v_h  < 100\kms$ are  plotted as filled
circles, stars with $100\kms <  v_h < 170\kms$ displayed with ``star''
symbols, stars with $170\kms < v_h < 380\kms$ are plotted as dots. The
two remaining stars with $v_h > 400\kms$ are marked as open circles.
\label{figCMD}}
\end{figure}

\begin{figure}
\centerline{ \psfig{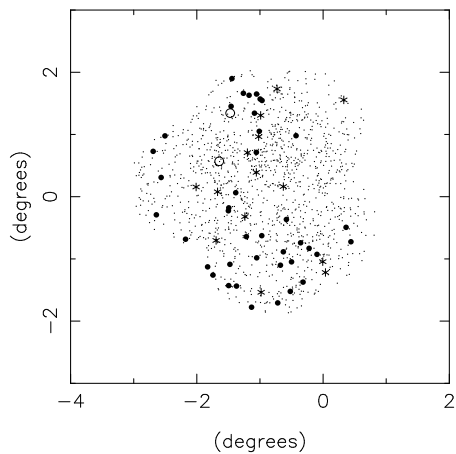}}
\caption[]{The  location of  the target  objects in  the  LMC.  The
stars are marked according to their radial velocities, as in Figure~1.
\label{figLoc}}
\end{figure}

\begin{figure}
\centerline{ \psfig{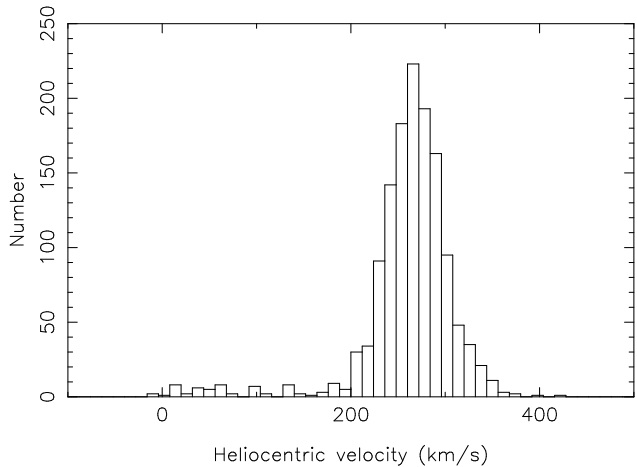}}
\caption[]{The heliocentric  radial velocity distribution  of stars in
the sample.
\label{figVrhist}}
\end{figure}

\section{Observations}

A sample  of $\sim 6000$ stars  was selected from an  LMC Color Magnitude
Diagram (CMD) produced
from  UKST  photographic plates  measured  on  the  APM facility  (see
Figure~\ref{figCMD}).   These   stars  were  randomly   selected  with
magnitudes  and  colours  in  the ranges  $16<R<18$  and  $-1.0<B_{\rm
J}-R<2.5$, in order  to have as few biases as  possible on the sample
selection.   The  fields  were  chosen  centered  on  each  previously
microlensed star.

The  present dataset  is  the  result of  3  observing campaigns  from
1999-2002,  using the  2dF (2  Degree  Field) instrument  at the  3.9m
Anglo-Australian  Telescope. This instrument  is a  fibre spectrograph
coupled to a robot positioner that is capable of observing the spectra
of up to 400 sources simultaneously over a 2-degree diameter circular
field.  Due  to adverse  weather conditions only  5 field  setups were
obtained, all  in UK Schmidt Telescope  field 56, for a  total of 1576
target stars observed.   Figure~\ref{figLoc} shows the distribution of
these sources over the face of the LMC.  Table~1 provides a summary of
the observing  log.  The 1200V  grating was used, giving  spectra with
dispersion  of  1.1\AA/pixel  over  the wavelength  range  4625\AA$  <
\lambda <  5765$\AA.  For each  fibre setup, three  1800~sec exposures
were combined.

\begin{table}
\begin{center}
\caption{Summary of Observations}
\begin{tabular}{lrrrr} \hline \hline
Date & Field & RA & DEC & Epoch \\ \hline
Nov 24 2000 & F056 Conf 05 & 05 04 03  &   -69 33 18  &   2000\\
Jan  5 2002 & F056 Conf 01 & 05 14 44  &   -68 48 01  &   2000\\
Jan  5 2002 & F056 Conf 02 & 05 17 14  &   -70 46 58  &   2000\\
Jan  5 2002 & F056 Conf 04 & 05 26 14  &   -70 21 14  &   2000\\
Jan  5 2002 & F056 Conf 21 & 05 24 03  &   -68 49 12  &   2000\\  \hline
\end{tabular}
\end{center}
\end{table}

The  spectral  images  were  debiased,  flat-fielded  and  wavelength
calibrated using the excellent 2dF pipeline software.  The spectrum of
the night sky was monitored with $\simgt 20$ fibres in each setup, and
mean  sky spectrum was  subtracted from  the object  spectra.  Judging
from  the residuals  at the  position of  the bright  [OI]5577\AA\ sky
line, the  sky subtraction  process is accurate  to better  than $\sim
1$\%.  After combining  the 3  exposures, the  average quality  of the
spectra  of  our  survey stars  was  greater  than  $S/N =  10$.   The
wavelength  calibration  was  checked  by  fitting  the  peak  of  the
[OI]5577\AA\ sky  line.  We estimate that 
the typical velocity error is $\sim 15\pm 10\kms$.

Due to the wide colour range of the stars observed (${\rm -1 < B_J-R <
2.5}$), the cost  of observing a sufficient number  of radial velocity
standard stars of spectral types covering this colour range would have
been prohibitive  with 2dF.  Instead we  chose to use  survey stars as
radial velocity standards,  picking 10 high S/N stars  at roughly even
intervals of ${\rm  B_J-R}$ colour over the above  range.  Each survey
star was crosscorrelated against each template star, giving a relative
velocity measurement,  the corresponding uncertainty,  and Tonry-Davis
cross-correlation   ``R''  parameter   \citep{tonry79}.   The  template
spectrum that gave  the highest ``R'' value was  considered to provide
the best crosscorrelation match with a particular survey star, and the
resulting  radial   velocity  difference  was  used   to  compute  the
heliocentric  radial  velocity of  the  survey  star.  The  zero-point
velocity of one of the  (K-type) LMC radial velocity ``standards'' was
measured with  respect to  a genuine radial  velocity standard  of the
same spectral  type.  The radial  velocity offsets of the  remaining 9
LMC radial velocity ``standards''  were calculated by fitting the peak
of  the numerous  LMC population  (where we  assume that  there  is no
systematic  offset in  velocity  between stars  of different  spectral
types in the LMC).

\begin{figure}
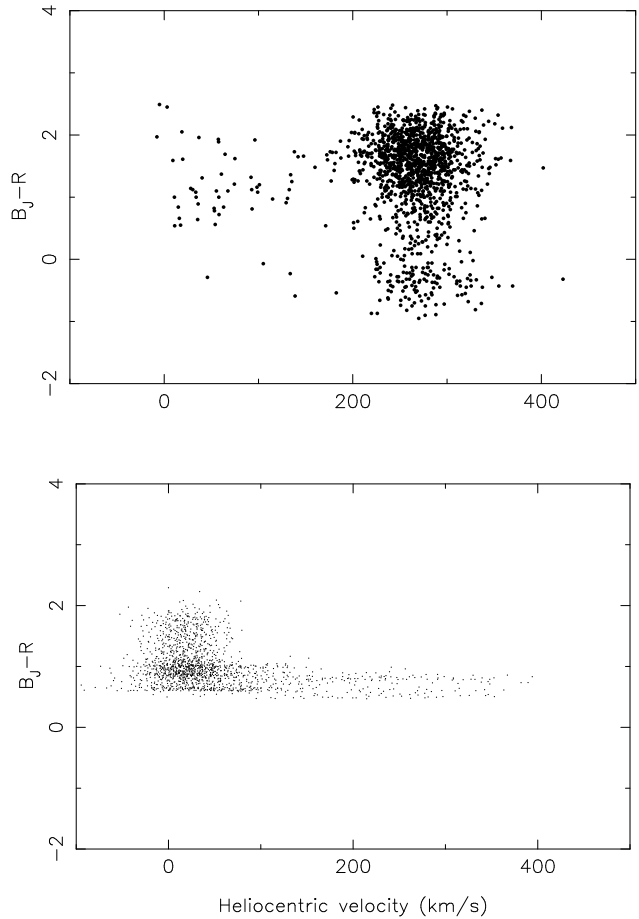

\centerline{ \psfig{figure=MC964fig4a.eps,angle=270,width=3.4in}}
\centerline{ \psfig{figure=MC964fig4b.eps,angle=270,width=3.4in}}
\caption[]{Heliocentric radial velocity as  a function of star colour.
Upper panel: data. Lower Panel: predictions of the Galaxy model
with a color conversion ${\rm Bj-R = 1.25(B-V)}$.  Note
the outliers in the data cannot be explained by the Galaxy model.
\label{figVr-color}
\label{figGalVr-color-mod}}
\end{figure}

To ensure  a good quality sample,  we applied three  parameter cuts to
the dataset: we require the Tonry-Davis cross-correlation parameter to
be $R>5$;  the heliocentric radial velocities are  constrained to have
$|v_{h}|   <  1000\kms$;   and  the   radial   velocity  uncertainties
(determined from the fit of the cross-correlation peak) were set to be
$\delta v_{h} < 50\kms$. The  resulting dataset, cleaned of low signal
to noise spectra and spectra that were not well matched by our library
of  10  ``standard''  stars,  comprises 1347  objects.   The  velocity
histogram of  this sample  is displayed in  Figure~\ref{figVrhist}.  A
scatter  diagram is  also  shown in  Figure~\ref{figVr-color} for  the
velocity and the color of our sample stars.

\section{The expected Galactic and LMC populations}

The  LMC is  located  at $\ell=280^\circ$,  i.e.   roughly toward  the
direction of  anti-rotation, and is  below the disk  at $b=-33^\circ$.
Since the Sun  rotates about the Galactic Centre  $10\kms$ faster than
the LSR \citep{dehnen98}, most local  stars will be seen towards lower
heliocentric radial velocities.  However,  the main effect is that the
projection  of the  circular  velocity vector  of  the disk  decreases
beyond the  tangent point.  Thus  the majority of Galactic  disk stars
seen along the line of sight towards  the LMC will be seen with $v_h <
0\kms$.     In    Figures~\ref{figGalVr-color-mod}(lower-panel)    and
\ref{figGalVr}  we  display   a  kinematic  star-counts  model,  which
contains  disk, thick  disk  and spheroid  populations.  The  expected
number of sources per square degree within our magnitude range is 1012
(disk:  dashed),  620  (thick  disk: dot-dashed)  and  370  (spheroid:
dotted) \citep{ibata94}.

Note that, of  the stars with $v_h <  100\kms$, most have intermediate
colours   ${\rm   B-V   \sim   1}$,  and   are   therefore   typically
Solar-luminosity    main-sequence   stars    with    $M_R   \sim    6$
\cite[e.g.][]{binney98}.   The  distance  range corresponding  to  our
magnitude selection  range $16 < R  < 18$, is  then $\sim 1\kpc <  d <
2.5\kpc$. Given the Galactic latitude  of the fields, these disk stars
are located  several scale heights  below the disk, and  mostly beyond
the  tangent point.   Nearby, intrinsically  fainter, and  redder disk
dwarfs hardly make it into the magnitude selection region.

The simplest explanation of our data is as follows:

\begin{figure}
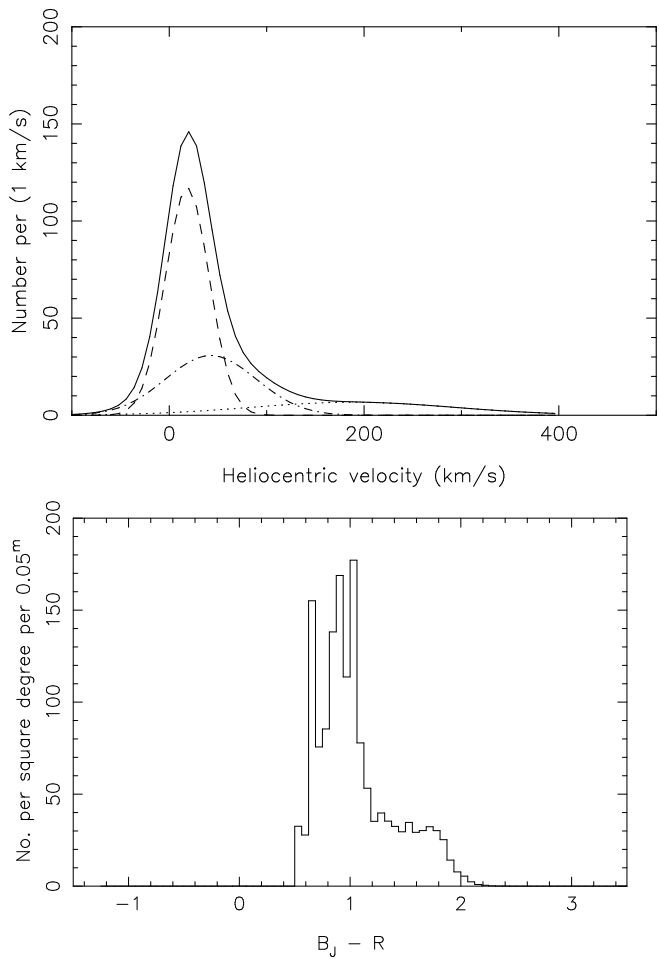

\centerline{ \psfig{figure=MC964fig5a.eps,angle=270,width=3.4in}}
\centerline{ \psfig{figure=MC964fig5b.eps,angle=270,width=3.4in}}
\caption[]{
The  expected  velocity  distribution
(upper panel) and color distribution (lower panel) of  Galactic  stars,
according to  the Galactic  kinematics model of  \citet{ibata94}.  The
velocity contributions of  the  disk,  thick  disk  and  stellar  spheroid  are
displayed,  respectively,  by a  dashed,  a  dot-dashed  and a  dotted
line. The solid line shows the sum of the three components.
\label{figGalVr}\label{figGalcolor}}
\end{figure}

\begin{enumerate}
\item
The 40 stars with $v_h <  100\kms$ are normal disk dwarfs, and display
an expected radial velocity distribution.
\item
The 1291 stars with $170\kms <  v_h < 380\kms$ are drawn from a single
simple    Gaussian    distribution,    with    intrinsic    dispersion
$\sigma_v=24\kms$.
\item
The 3 stars with $100\kms < v_h  < 170\kms$ and ${\rm B_J-R < 0}$ have
suspect  velocities because  their radial  velocities as  derived from
different (but similar) templates do not agree.
\item
The 11  stars with $100\kms <  v < 170\kms$  and ${\rm Bj-R >  0}$ are
spheroid stars.
\item
The 2 stars with $v > 380\kms$ are halo stars.
\end{enumerate}

According to the model  displayed in Figure~\ref{figGalVr}, the number
of stars in the velocity range $100\kms < v_h < 170\kms$ should number
only  5\% of  the  number in  the  velocity range  $-100\kms  < v_h  <
100\kms$.  The  observed number is 35\%.   Is the excess  of $\sim 11$
stars due  to an extraneous  unexpected population that  could account
for the observed microlensing?  This population, however, would only
represent a  small fraction  of the mass  required to account  for the
observed frequency microlensing. Hence this data effectively rules out
the existence  of an extraneous kinematic population  that exceeds 1\%
of the LMC.

\section{Comparison with previous work}

\begin{figure}
\centerline{ \psfig{figure=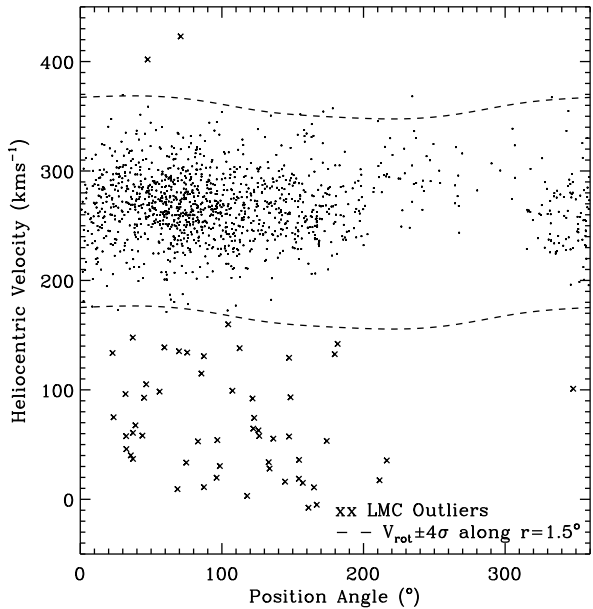,angle=0,width=3.4in}}
\caption[]{A scatter  plot of the  observed velocity vs.  the position
angle (from the North and  centered on the LMC center).  Large symbols
denote velocity  outliers.  The two dashed sinusoidal  curves show the
projected rotation pattern  of the LMC disk (predicted  from models of
van der Marel  et al. 2002 at $r=1.5^\circ$ from  the LMC center), but
offset by $4\sigma$  where  $\sigma \sim  24\kms$  is the  velocity
dispersion of  our LMC  sample).  Note that  our sample fields  do not
cover all PA's uniformly, and  our fields are mostly on the short-axis
(PA$=129.9^\circ$) of the LMC disk.
\label{figVr-PA}}
\end{figure}

The distribution of radial velocities of LMC stars has been studied by
many authors.  The kinematical properties of the LMC have been studied
with  many different  tracers, including  HI  \cite[ e.g.]{kim98}, star
clusters    (\citealt{freeman83,  schommer92}),    planetary   nebulae
\citep{meatheringham88},      HII      regions     and      supergiants
\citep{feitzinger77}.   Recently \citet{marel02} 
have combined  the carbon  stars dataset  of \citet{kunkel97}.
which covers the periphery of the LMC at all position angles, with 573
stars  from \citet{hardy02},  which were  selected from  the  surveys of
\citet{blanco80}.  \citet{marel02} describe  the radial velocity field
with a  rotation of stars  in the LMC  centered on $\alpha_{\rm  CM} =
5^{h} 27.6^{m} \pm 3.9^{m}$ and $\delta_{\rm CM} = -69^{\circ} 52' \pm
25'$, which is  about $1.2^{\circ} \pm 0.6^{\circ}$ away  from the gas
kinematical center.  The kinematics  of stars are roughly described by
a  flat rotation  of $50\kms$  outside $4^\circ$  with  an inclination
$i=34.7^\circ\pm      6.2^\circ$     and     a      position     angle
$\Theta=129.9^\circ\pm  6^\circ$.   The  velocity dispersion  for  the
carbon  stars is 20--$22  \kms$ between  $1$ and  $3.5 \kpc$  from the
center, followed  by a decline to  16--$17 \kms$ between  $3.5$ and $7
\kpc$  from the  center, and  a subsequent  increase to  21--$22 \kms$
between $7$ and $9 \kpc$.  The velocity dispersion increases with age
\citep[ e.g.]{gyuk00}, with a range  from $\sigma \approx 6
\kms$ for the youngest populations (e.g., supergiants, HII regions, HI
gas) to $\sigma \approx 30 \kms$ for the oldest populations (e.g., old
long-period variables,  old clusters).  The  carbon stars are  part of
the  intermediate-age  population  which  is  believed  to  be  fairly
representative for the bulk of the mass in the LMC.  Our sample appear
to  be consistent  with  the carbon  star  sample of  \citet{marel02}.
Figure~\ref{figVr-PA} is made to examine the question whether there is
any significant  velocity variation with spatial  position.  Given the
sparseness  of  our  sample,  we  do not  see  any  strong  systematic
variation  across our  fields.   Our  fields are  also  mostly on  the
short-axis of  the LMC  disk, hence the  signature due to  rotation is
also very weak.  This supports our treatment of our sample as a whole.

\section{Analysis}

We now  investigate the question of  whether we could  have detected a
significant population superimposed on the LMC.

Figure~\ref{figGalVr-color-mod}  is  a  diagram  similar to  the  Hess
diagram, but shows the Galactic model projected in the radial velocity
and  colour  plane.   We  normalise  the Galactic  model  using  those
neighbouring field histograms.  Compare  this with the distribution of
possible Galactic stars in our sample e.g. the 4 blue stars with $50 <
v  < 150$  km/s.  It  is difficult  for a  standard Galactic  model to
account for  the velocity distribution  we are observing even  for the
redder objects in the $0 < v < 150$ km/s range.

\begin{figure}
\centerline{ \psfig{figure=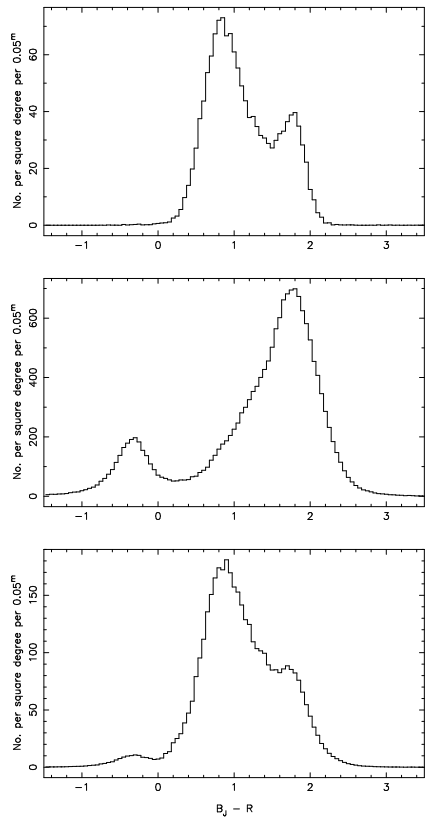,angle=0,width=\hsize}}
\caption[]{From top  to bottom color histograms for  UKST fields \#54,
\#56, and \#58.
Field    \#58    is   toward    the    Galactic    Plane   at    about
$(l,b)=(281^\circ,-25^\circ)$ and \#54 at $(286^\circ,-42^\circ)$ with
$E(B-V) = 0.11$ and $0.04$  respectively.  The middle panel
is for the reference field \#56, which
is roughly centered on the LMC at $(281^\circ,-34^\circ)$ 
with $E(B-V) = 0.23$.  We used a colour cut
which removes the stars in the blue and late M in the red.
\label{figGalcolorhist}}
\end{figure}

We  also  test  the Galactic  model  by  comparing  it with  a  colour
histogram from  our plate data  in that  region in the  $16 < R  < 18$
magnitude  range.   With  plausible  assumptions  about  completeness,
Figure~\ref{figGalcolorhist} shows the color  histograms in a few UKST
fields \#54, \#56  and \#58.  
Except for the reference field \#56, which is on the LMC,
these fields are  sufficiently away from
the  LMC  to  represent  the  Galactic  distribution  in  the  general
direction of the LMC.
While there are significant differences between \#54 and \#58,
the overall distribution resembles the prediction from our Galaxy model 
(cf. Fig.~\ref{figGalcolor}), i.e., most of Galactic stars 
in the LMC direction have a largely bimodal color distribution
peaked around ${\rm Bj-R = 1.25(B-V)=1}$ and $1.5$.
Our Galaxy model, albeit simple, should not significantly bias our conclusion.

\section{Conclusions}

If the  microlensing in  the LMC is  indeed due to  self-lensing, then
kinematic outliers are expected towards the LMC in their distributions
of radial velocity, proper  motion, projected distance from the center
of the LMC, distance modulus and reddening (Zhao 1999a,b).  The lack
of features  in our  radial velocity sample  argues tentatively  for a
relatively smooth  and uniform distribution  of stars in the  LMC with
very few  outliers.  Our data does  not support the notion  that there
is a large fluffy stellar halo around the LMC (e.g., Evans \& Kerins 2000).

On the  other hand,  it is very  unlikely for the  hypothetical (Milky
Way)  halo   white  dwarfs   to  account  for   the  lensing   in  the
LMC. \citet{zhao02}  argues that any violent  galactic winds following
early epoch  of star bursts  would significantly weaken  the potential
wells of galaxies.  The Milky Way galaxy would have been disintegrated
if more than half of its  dynamical mass were blown off violently in a
wind generated by the  formation of numerous macho progenitors.  Since
the Milky  Way galaxy is very tightly  bound, and there is  no sign of
any globular cluster escaping from  the Milky Way either, these should
imply  an upper  limit on  the  baryons participating  the early  star
bursts and baryons  locked in stellar remnants, such  as white dwarfs.
The white dwarfs should not make  up more than 1-5\% of the total mass
of  the Galaxy.   Similar arguments  also imply  upper limits  for the
amount  of neutron  stars and  stellar black  holes, in  galaxy halos.
This dynamical  upper limit  is exceeded by  the amount of  halo white
dwarfs  claimed  in  recent  proper  motion searches  and  in  earlier
microlensing observations (Alcock et  al.  2000) in the Galactic halo,
suggesting  these interpretations  of observations  to  be problematic
theoretically.

We propose that the most likely one of existing models for
microlensing in the LMC is perhaps the ``mutual-lensing'' models of
\citet{zhao01}, where the LMC stars are on two distinct planes with
different inclinations or warped
\footnote{The LMC disk is known to have a poorly defined inclination 
observationally, and is perhaps warped (e.g., Olsen \& Salyk 2002)} 
so that stars on the front plane could lense
those in the plane $\sim 1$kpc behind.  One of the planes may be the LMC disk,
and the other may well be 
the irregular bar of the LMC, which is known to be offset from the 
LMC disk.  Such a unvirialised configuration could only be a transient phase 
of the general process of merging, after perhaps
recent interactions with SMC and the Milky Way.
In this type of models we do not expect the stars in the two planes
to share the same rotation around the LMC disk, 
but the differences in systematic velocities
should be small because they share a common potential well.
Our data can exclude any systematic offset of $\ge 70\kms$ safely, 
but not yet exclude the possibility of smaller velocity
distortion due to different inclinations in the LMC.  
A larger sample with more accurate
velocity would be needed to reveal small differences among disk or bar
stars of the LMC, and set stronger limits on star-star lensing in the LMC.

We thank Brian Boyle for undertaking the service observing at the AAT.  

\newcommand{\aap}{A\&A}
\newcommand{\apj}{ApJ}
\newcommand{\apjl}{ApJ}
\newcommand{\aj}{AJ}
\newcommand{\mnras}{MNRAS}
\newcommand{\apss}{Ap\&SS}
\newcommand{\nat}{Nature}

\end{document}